\documentclass[seceq]{ptptex}

\usepackage[final]{epsfig}

\usepackage{wrapft}

\pubinfo{}%Editorial Office will fill
\title{%        %You can use \\ for explicit line-break.
Jets in medium
}
\subtitle{what RHIC and LHC measurements of $R_{AA}$ and $I_{AA}$ can teach about the parton-medium interaction}    %Use this when you want a subtitle.

\author{%       %Use \scshape  for the family name.
Thorsten \textsc{Renk}$^{1,2}$\footnote{email: thorsten.i.renk@jyu.fi} 

}

\inst{%         %Affiliation, neglected when [addenda] or [errata].
$^1$Department of Physics, P.O. Box 35, FI-40014 University of 
             Jyv\"askyl\"a, Finland\\
$^2$Helsinki Institute of Physics, P.O. Box 64, FI-00014 University 
             of Helsinki, Finland
}

\recdate{%      %Editorial Office will fill in this.
}

\abst{%         %This abstract is neglected when [addenda] or [errata].
High $P_T$ observables associated with hard pQCD processes are regarded as tomographic probes of the medium created in ultrarelativistic heavy-ion collisions. Unfortunately, tomo\-graphy is unexpectedly complicated in practice due to model uncertainties and the highly averaged nature of many observables. However, a systematic comparison study demanding a simultaneous description of many observables within the same framework is sufficiently powerful to rule out a large fraction of conjectured models about the physics mechanism of parton-medium interaction. From this study, it can be inferred that the relevant mechanism is likely to be perturbatively tractable medium-induced radiation with a small component of elastic energy transfer into the medium.
}

\begin{document}
\maketitle

\section{Introduction}

High-$P_T$ processes taking place in the background of the medium produced in ultrarelativistic heavy-ion (A-A) collisions are a cornerstone of the experimental A-A program at the LHC. The aim is to use these processes to do ``jet tomography'', i.e. to study both the short-distance physics of 
the bulk medium (i.e. its relevant degrees of freedom) and long-distance physics, i.e. the geometry of the medium expansion. This includes quantifying various transport coefficients in the medium ($\hat{q}, \hat{e}$,\dots) as well as constraining properties of the initial state geometry (CGC vs. Glauber profile, size scale of fluctuations, \dots). Since hard parton scattering processes take place (by virtue of the uncertainty relation) before a medium can be formed, their rate is perturbatively calculabe as in vacuum, however the resulting evolution via a parton shower to a hadronic jet is modified by the final state interaction with the medium. The obvious strategy to get tomographic information about the medium is thus to compare fully reconstructed jets in vacuum to jets in medium.

In vacuum, jets are defined by algorithms combining hadrons or calorimeter towers in a certain way. The algorithms are cleverly chosen to undo the perturbative Quantum Choromodynamics (pQCD) evolution of the shower, such that jets on the detector level (calorimeter towers) are approximately equal to jets on the hadron level (particle tracks) and jets on the parton level (the output of pQCD calculations). This works because the flow of energy and momentum must always remain inside the jet. However, that is no longer true in a medium, as the perturbatively evolving shower may exchange energy and momentum with the non-perturbative medium. As a result, there is no unambiguous notion what a jet in medium should be: If one defines the perturbative part of the shower as jet, then the energy of jets will be smaller than the energy of the original parton, as part of its energy may be carried by non-perturbative excitations of the medium. If a jet is defined as everything causally correlated with the shower-initiating parton, then the jet energy will be in general higher than the parton energy, as collisions with medium partons may correlate them with the jet, while part of their energy is thermal. Finally, if the jet is taken to be the flow of energy and momentum of the original parton, then the jet energy will be the parton energy, but a jet can no longer be defined at the hadron level. 

Complications like these, having to do with the need to separate jet and medium physics, make it appealing to look at purely perturbative objects first, i.e. the leading hadrons of a jet. These are accessible in single inclusive high $P_T$ spectra in the nuclear suppression factor $R_{AA}$ and in the suppression factor of correlations $I_{AA}$. There is some systematics available in $\sqrt{s}, P_T$, centrality or the angle $\phi$ with the reaction plane, and the strategy of a systematic analysis is to deduce the nature of the parton-medium interaction from these observables.

\section{Pathlength and centrality dependence}

In the following, we test different combinations of medium evolution and parton-medium interaction models against a large body of high $P_T$ observables. In particular, for the medium evolution we use a 3+1d ideal \cite{hyd3d}, a 2+1d ideal \cite{hyd2d,hydEbyE}
and a 2+1d viscous hydro code \cite{vhyd} with both CGC and Glauber initial conditions. On the parton-medium interaction side, we test a radiative energy loss model \cite{ASW}, a parametrized \cite{Elastic} and a Monte-Carlo (MC) model \cite{ElasticMC} for incoherent energy loss, a strong-coupling phenomenological model based on AdS/CFT ideas \cite{AdS} and the MC in-medium shower code YaJEM \cite{YaJEM1,YaJEM2} with its variant YaJEM-D \cite{YaJEM-D} which introduces an explicit pathlength/energy dependence into the minimum virtuality scale down to which the shower is evolved.
A suitable criterion to distinguish models is the pathlength dependence of the medium effect, and the relevant observable is $R_{AA}(\phi)$.
The first task is to constrain combinations of medium and parton-medium interaction model and to quantify how much the medium geometry influences the results. This has been done in \cite{jet-hyd-sys}. The main finding is that any pathlength dependence which is effectively linear can be constrained to be less than $\sim 10$\% of the total effect. This rules out a large contribution of either elastic energy loss \cite{Elastic,ElasticMC} or radiative energy loss with finite energy corrections \cite{YaJEM1,YaJEM2}.

\section{$\sqrt{s}$ and $P_T$ dependence}

The $P_T$ dependence of single high $P_T$ hadron suppression arises from a combination of the $p_T$ dependence of the primary parton spectrum and the probability distribution of small energy losses $P(\Delta E)$ for the leading shower parton \cite{JyvProc}, in addition to any other explicitly $P_T$ dependence in the parton-medium interaction model. At RHIC, kinematic restrictions do not allow to probe a large range of $P(\Delta E)$. This, however, is different at the LHC. An extrapolation of RHIC fits to $R_{AA}$ data using a \emph{controlled} extrapolation of the medium from RHIC to LHC should therefore discriminate between different models based on how $P(\Delta E)$ looks in the vicinity of zero and if there is an explicit $p_T$ dependence in the suppression. This has been studied in \cite{RAA_LHC}. The advantage of this study is that the $P_T$ dependence of $R_{AA}$ is not very sensitive to the medium geometry.

\begin{figure}[htb]
\epsfig{file=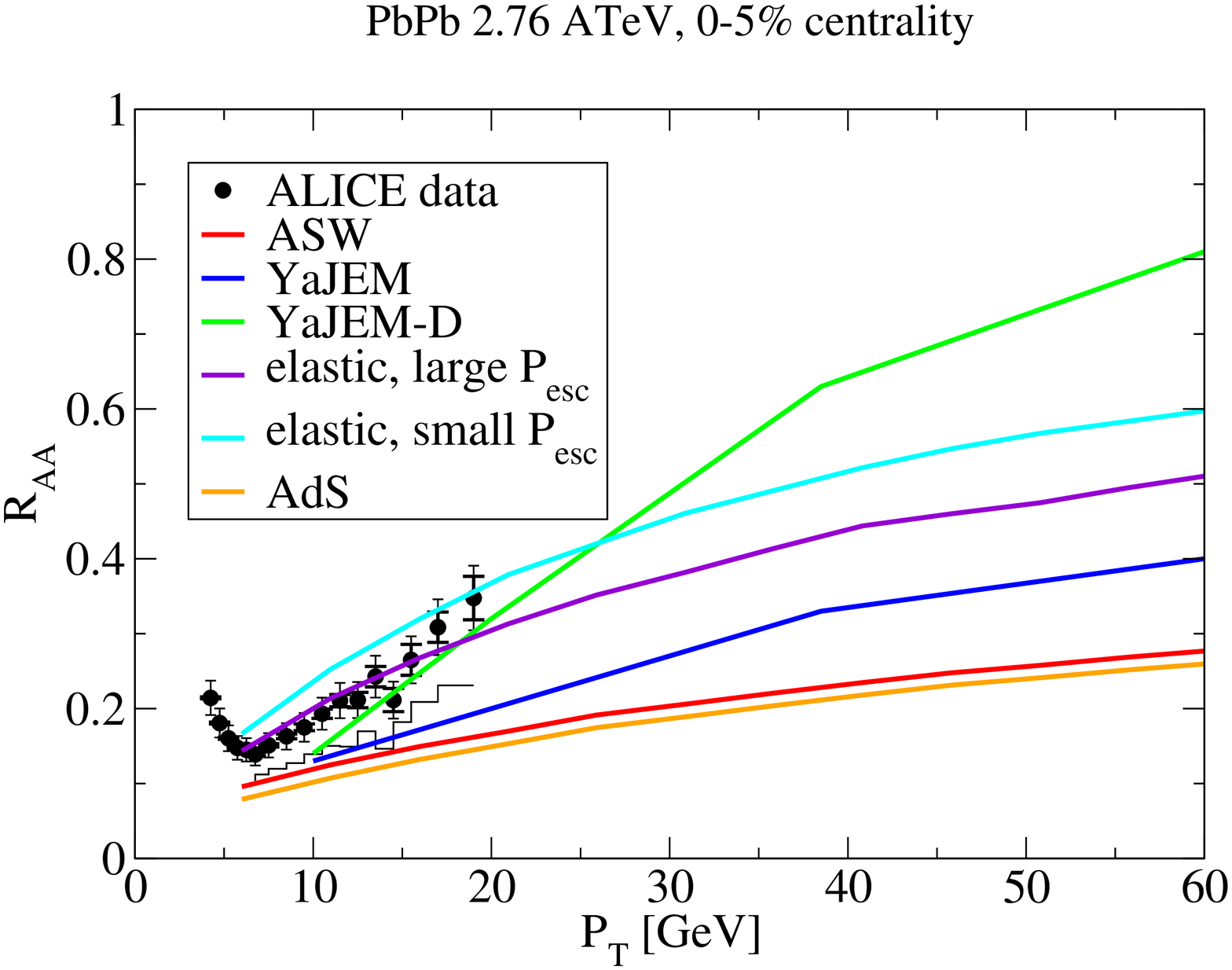, width=4.6cm}\epsfig{file=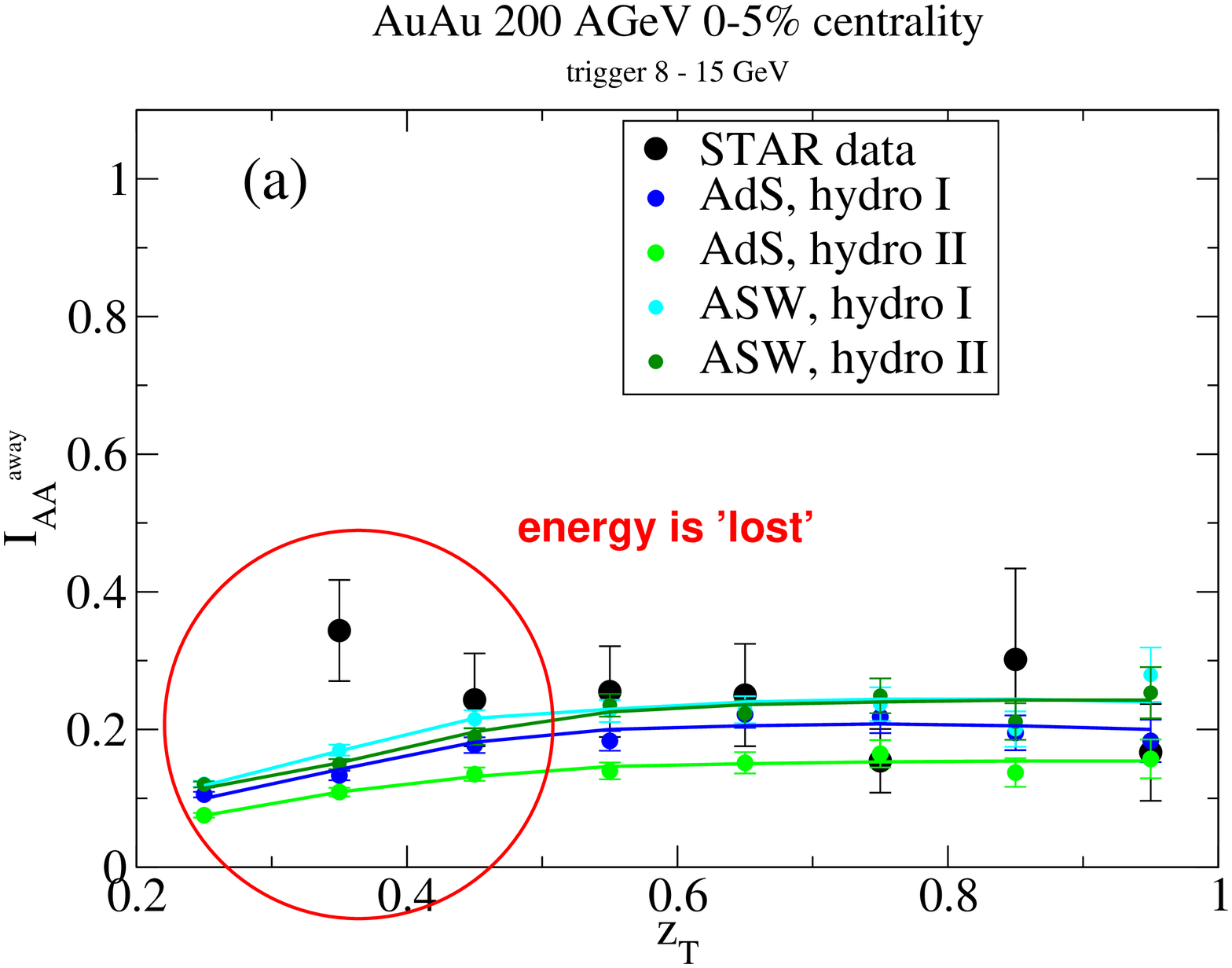, width=4.6cm}\epsfig{file=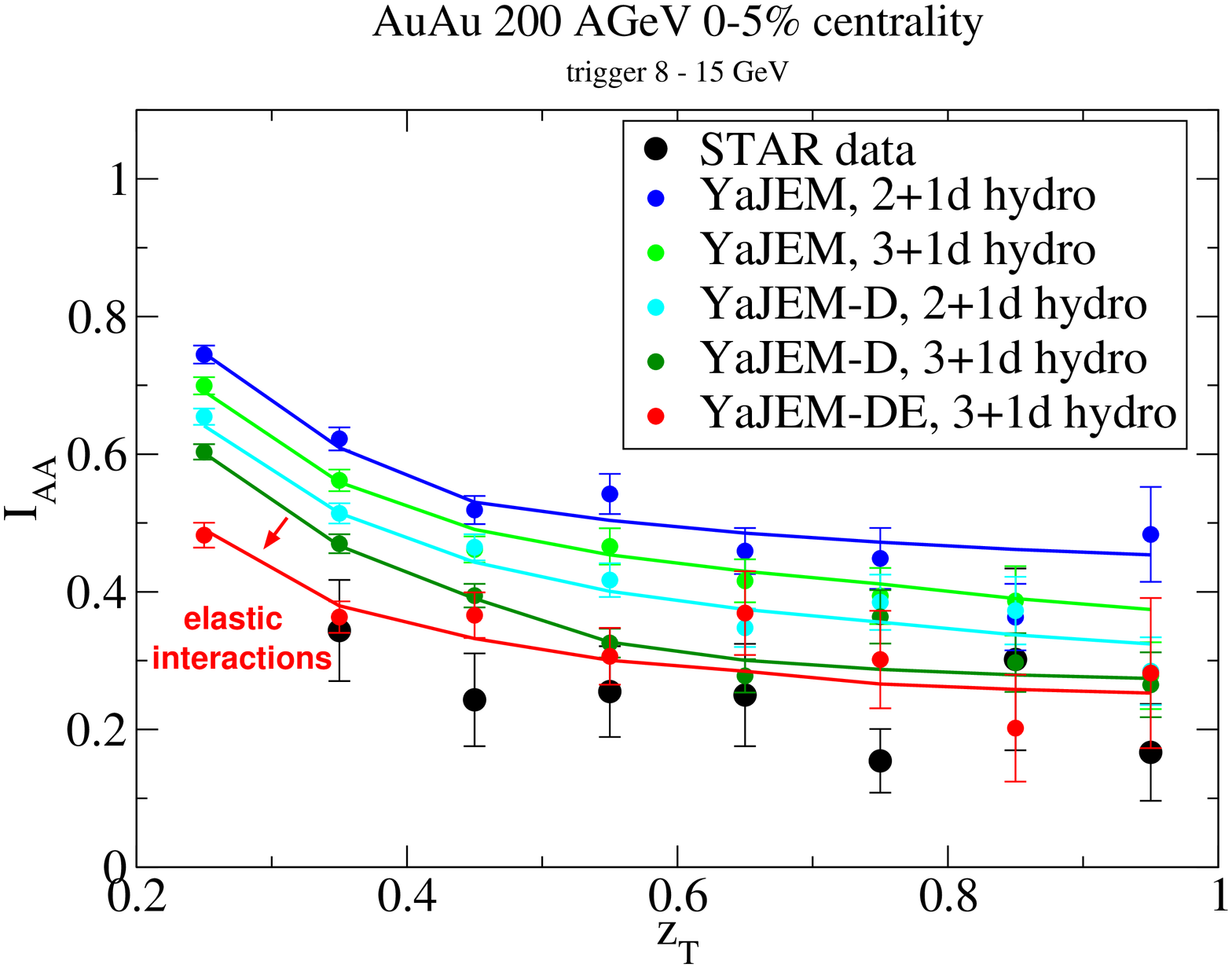, width=4.6cm}
\caption{\label{F-1}
Left panel: Nuclear suppression factor $R_{AA}$ as measured by the ALICE collaboration in central 2.76 ATeV Pb-Pb collisions \cite{ALICE}, compared with various model calculations (see text). Central Panel: Suppression factor $I_{AA}$ of the away side correlation strength measured in central 200 AGeV Au-Au collisions \cite{STAR} compared with leading particle energy loss models (see text). Right panel: same as central panel, but compared with in-medium shower evolution models.
}
\end{figure}

The main findings are shown in Fig.~\ref{F-1} (left) for the models mentioned above. They can be quantified by a parameter $R$ which determines the relative amount of refit of model parameters from RHIC to LHC needed for the best possible description of LHC data (where $R=1$ indicates a perfect extrapolation from RHIC to LHC). The value of $R$ for various models is found in Tab.~\ref{T-1}.

\begin{table}
\begin{center}
\begin{tabular}{|c|c|c|c|c|}
\hline
&YaJEM-D & YaJEM & ASW & AdS\\
\hline
R & 0.92 & 0.61 & 0.47 &  0.31\\
\hline
\end{tabular}\label{T-1}
\end{center}
\end{table}

It is evident that the model based on strong coupling dynamics (AdS) has the worst extrapolation. This is directly connected with the success of the same model for pathlength dependent observables: Since it generates energy loss $\sim L^3$, for dimensional reasons it probes the medium temperature with $T^4$ instead of $T^3$ as in all other models. It is this difference in the temperature extrapolation from RHIC to LHC which makes AdS fail more badly than other models. Note that the only viable model after pathlength and $\sqrt{s},P_T$ dependence have been discussed is YaJEM-D which has an explicit energy dependence in the parton-medium interaction.

\section{Dihadron correlations}

We may proceed to test the suppression of away side triggered correlation pattern systematically in terms of $I_{AA}$. This is a more differential observable based on a conditional probability, which implies a number of interesting biases. This has been done in \cite{IAA} and the main results are shown in Fig.~\ref{F-1} (middle and right panel).

It is evident that all models making an energy loss approximation for the leading parton instead of computing the full in-medium shower (middle panel) fail even qualitatively in the description of the low $z_T$ region where subleading shower hadrons are relevant. We may take this as evidence that medium-induced gluon radiation is indeed observed and can be used to constrain models.
As apparent from the right panel, YaJEM-D reproduces qualitatively the right behaviour. However, only of this model is supplemented by 10\% elastic energy loss \cite{IAA_elastic} (the maximum allowed by pathlength dependence studies) is quantitative agreement with the data reached.

\section{Conclusions}

We can infer from these studies that a pQCD picture of medium-induced radiative energy loss and a small elastic component is compatible with the data whereas more exotic scenarios like a strong coupling description (AdS) are not. The amount of elastic energy loss can be constrained from above and from below to about 10\%, leading to the conclusion that the scattering partners probed by a hard parton are not light free quasiparticles (which would lead to about 50\% elastic contribution \cite{ElasticMC}).

In the most successful model description, full jets are expected to appear largely unmodified in the perturbative momentum regime except for an overall shift in energy \cite{Jets}, which qualitatively agrees with experimental results by CMS \cite{CMS-dijets}. This gives rise to the expectation that qualitatively the physics of parton-medium interaction is now known and a precision analysis with the aim of extracting tomographic information can be the next step.

\section*{Acknowledgements}
This work was supported by the Finnish Academy (projects 130472 and 133005).

\end{document}